\documentstyle[12pt,epsfig]{article}

\begin{document}

\newpage
\setcounter{page}{0}

\begin{titlepage}
\begin{flushright}
\hfill{YUMS 98--06}\\
\hfill{SNUTP 98--30}\\

\hfill{\today}
\end{flushright}
\vspace*{1.0cm}

\begin{center}
{\large\bf CP Violation in the Cabbibo-suppressed Decay 
            $\tau\rightarrow K\pi\nu_\tau$\\
            with Polarized $\tau$ Leptons}
\end{center}
\vskip 1.cm
\begin{center}
{\sc S.Y.~Choi$^{\mathrm{a}}$, Jake Lee$^{\mathrm{a}}$ and 
     J.~Song$^{\mathrm{b}}$}

\vskip 0.8cm

\begin{small} 
$^{\mathrm{a}}$ Department of Physics, Yonsei University 120-749, Seoul, 
                  Korea. \\
\vskip 0.2cm
$^{\mathrm{b}}$   Center for Theoretical Physics, Seoul National University, 
                  Seoul 151-742, Korea.
\end{small}
\end{center}

\vskip 2cm

\setcounter{footnote}{0}
\begin{abstract}
CP violation from physics beyond the Standard Model (SM) is investigated in the 
Cabbibo-suppressed decay $\tau\rightarrow K\pi\nu_{\tau}$ with polarized $\tau$
leptons, to which both the ${\rm J}^{\rm P}=1^-$ resonance $K^*$ and the 
${\rm J}^{\rm P}=0^+$ resonance $K^*_0$ contribute. 
In addition to the CP-odd rate asymmetry, $\tau$ polarization enables us to
construct three additional CP-odd polarization asymmetries that can be enhanced 
due to the interference between the $K^*$ and $K^*_0$, and whose magnitudes 
depend crucially on the $K^*_0$ decay constant, $f_{K^*_0}$. 
Taking a QCD sum rule estimate of $f_{K^*_0}=45$ MeV and the present
experimental constraints on the CP-odd parameters into account, we estimate 
quantitatively the maximally-allowed values for the CP-odd rate and 
polarization asymmetries in the multi-Higgs-doublet (MHD) model and 
the scalar-leptoquark (SLQ) models consistent with the SM gauge symmetry 
where neutrinos are massless and left-handed as in the SM.
We find that the CP-odd rate and polarization asymmetries are of 
a similar size for highly-polarized $\tau$ leptons and, for their 
maximally-allowed values, CP violation in the MHD model and two SLQ models 
may be detected with about $10^6$ and $10^7$ $\tau$'s at the $1\sigma$ level.
\end{abstract}
\vskip 1cm
\centering{\bf (to be submitted to PLB) }
\end{titlepage}

\newpage
\renewcommand{\thefootnote}{\alph{footnote}}

The decay of the $\tau$, the most massive of the known leptons,
can serve not only as a useful tool in the investigation of some 
aspects of the SM but also as a powerful experimental probe of new physics 
phenomena \cite{Pich}. One phenomenon where new physics can play 
a crucial role is CP violation. In light of this aspect the $\tau$ decays 
into hadrons have been recently studied as probes of CP violation in the scalar 
sector of physics beyond the SM \cite{Nelson,Tsai,Mirkes,Choi}. 

In the present paper we extend the previous work \cite{Choi} by Choi, Hagiwara 
and Tanabashi to probe CP violation in the Cabbibo-suppressed 
$\tau$ decay $\tau\rightarrow K\pi\nu_\tau$ with polarized $\tau$'s.
The decay mode is dominated by the contributions 
of the two lowest vector and scalar resonances, $K^*$ and $K^*_0$, 
with different spins and relatively large width-to-mass ratios \cite{PDG96}:
\begin{eqnarray}
\begin{array}{clll}
 K^* : &{\rm J}^{\rm P}=1^-, &\ \  m_{K^*}=892\ \ {\rm  MeV},
       &\ \ \Gamma_{K^*}=50\ \ {\rm  MeV}\,,\\
 K^*_0 : & {\rm J}^{\rm P}=0^+, &\ \  m_{K^*_0}=1430\ \ {\rm MeV},
       &\ \ \Gamma_{K^*_0}= 287\ \ {\rm  MeV}\, ,
\end{array}
\end{eqnarray} 
and the mode is expected to have larger scalar contributions than the $2\pi$ or
$3\pi$ modes due to the $s$ quark mass much larger than the $d$ quark 
mass. In the light of these aspects, the Cabbibo-suppressed mode is worthwhile
to be investigated in detail.
  
Including possible contributions from new physics with massless left-handed
neutrinos, we can write the matrix element for the decay 
$\tau^-\rightarrow (K\pi)^-\nu_\tau$ in the general form 
\begin{eqnarray}
M=\sqrt{2}G_F\Bigl[(1+\chi)\bar{u}(k,-)\gamma^\mu P_- u(p,\sigma)J_\mu
                   +\eta\bar{u}(k,-)P_+ u(p,\sigma)J_S\Bigr],
\label{decaym}
\end{eqnarray}
where $P_\pm=(1\pm\gamma_5)/2$, $G_F$ is the Fermi constant, and $p$ and $k$ are 
the four momenta of the $\tau$ lepton and the tau neutrino, respectively. 
The parameters $\chi$ and $\eta$, which parametrize contributions from physics 
beyond the SM, are in general complex. In eq.~(\ref{decaym}), the helicity of 
the $\tau^-$ is denoted by $\sigma$ ($\sigma=\pm 1$) with its spin quantization 
direction along its neutrino momentum direction. The tau neutrino is 
assumed to be massless and left-handed as in the SM so that the helicity value 
is $-1/2$ as indicated by the negative sign in its spinor $u(k,-)$. 
The vector and scalar hadronic matrix elements 
\begin{eqnarray}
&& J_\mu= \sin\theta_C\langle (K\pi)^-|\bar{s}\gamma_\mu u|0\rangle,\nonumber\\ 
&& J_S= \sin\theta_C\langle (K\pi)^-|\bar{s}u|0\rangle,
\end{eqnarray}
with $\sin\theta_C=0.23$ for the Cabbibo angle $\theta_C$ are related through 
the Dirac equations to the $\bar{s}$ and $u$ quarks at the quark level and
their explicit form can be parametrized in terms of two form factors
$F_K(q^2)$ and $F_S(q^2)$:
\begin{eqnarray}
&&J_\mu=\sqrt{2}\sin\theta_C
  \left[F_K(q^2)\left(g_{\mu\nu}-\frac{q_\mu q_\nu}{q^2}\right)(q_1-q_2)^\nu
       +\frac{m^2_{K^*_0}}{q^2}C_KF_S(q^2)q_\mu\right], \nonumber\\
&&J_S=\sqrt{2}\sin\theta_C
    \left(\frac{m^2_{K^*_0}}{m_s-m_u}\right)C_KF_S(q^2),
\end{eqnarray}
where $q_1$ and $q_2$ are the four-momenta of $\pi$ and $K$, respectively,
$m_s$ and $m_u$ the $s$ and $u$ current quark  masses, and $q$ is 
the four-momentum of the $K\pi$ system; $q=q_1+q_2$.

The coupling strength $C_K$ denoting the scalar contributions is
determined by the $K^*_0$ decay constant $f_{K^*_0}$
and the coupling strength $g_{K^*_0K\pi}$ of the $K^*_0$ to $K\pi$
under the assumption of Br($K^*_0\rightarrow K\pi)=100\%$:
\begin{eqnarray}
C_K=\frac{f_{K^*_0}g_{K^*_0K\pi}}{\sqrt{3}m^2_{K^*_0}}.
\end{eqnarray}
The value of $g_{K^*_0K\pi}$ is $4.9\ \ {\rm GeV}$ from
the measured $K^*_0\rightarrow K\pi$ decay width $\Gamma(K^*_0\rightarrow 
K\pi)\approx 287\ \ {\rm MeV}$.
Even though the $K^*_0$ decay constant $f_{K^*_0}$ is not  
experimentally measured, it has been estimated by several model-dependent
methods. The QCD sum-rule estimate in the $K^*_0$ narrow-width approximation
is $f_{K^*_0}\approx 31$ MeV \cite{Narison} and an effective Lagrangian estimate 
of $f_{K^*_0}$ including the width effects leads to a larger value of about 
45 MeV \cite{Bramon}, which approaches the pole dominance result of 50 
MeV \cite{Ayala}. In light of these present crude estimates of the $K^*_0$ decay
constant, we adopt for the $K^*_0$ decay constant the effective Lagrangian
estimate \cite{Bramon}
\begin{eqnarray}
f_{K^*_0}=45 \ \ {\rm MeV},
\end{eqnarray}
anticipating that the decay constant $f_{K^*_0}$ is measured more precisely 
in future experiments.

\vskip 0.5cm

\begin{center}
\begin{figure*}[htb]
\hbox to\textwidth{\hss\epsfig{file=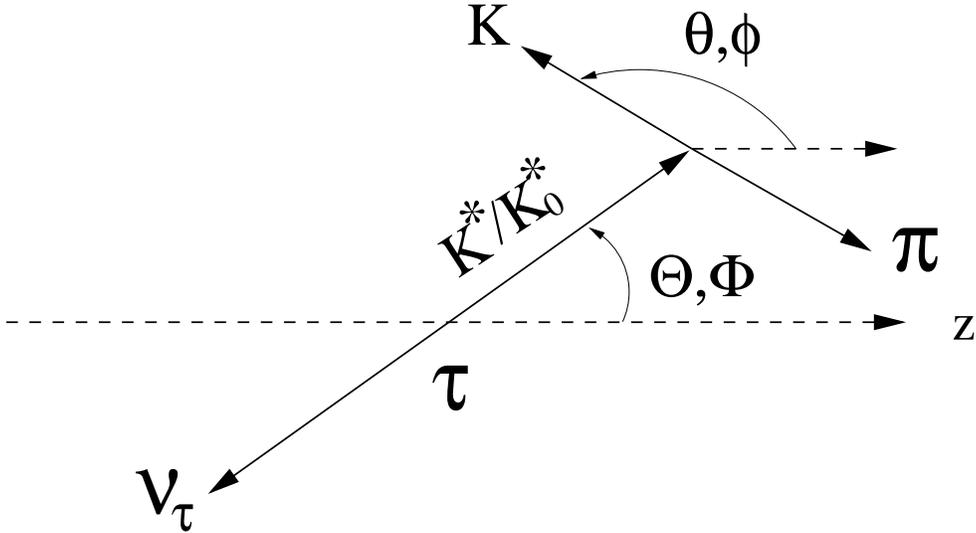,height=7cm}\hss}
% \vskip -2.cm
\caption{\it Definition of the angular variables in the $\tau$ and $(K\pi)$ 
             rest frames for the Cabbibo-suppressed $\tau$ decay 
             $\tau^-\rightarrow (K\pi)^-\nu_\tau$. 
             The $(K\pi)$ momentum direction in the $\tau$ rest frame 
             is denoted by the polar angle $\Theta$ and the azimuthal angle 
             $\Phi$ with respect to the $\tau$ momentum direction defined as 
             the positive $z$-axis, and the $K$ momentum direction in the 
             $(K\pi)$ rest frame denoted by the angles $\theta$ and $\phi$ 
             with respect to the same positive $z$ direction.}
\label{fig:decay}
\end{figure*}
\end{center}

In spite of the several resonance contributions to $F_K$ and $F_S$, the 
form factors are approximated to be the propagators of the lowest-level
resonance states $K^*$ and $K^*_0$: 
\begin{eqnarray}
F_K(q^2)=B_{K^*}(q^2), \qquad  
F_S(q^2)=B_{K^*_0}(q^2), 
\end{eqnarray}
respectively, where $B_{K^*}$ and $B_{K^*_0}$ are parametrized in the 
Breit-Wigner form with the momentum-dependent widths \cite{Tauola}:
\begin{eqnarray}
B_{X}(q^2)=\frac{m^2_X}{m^2_X-q^2-im_X\Gamma_X(q^2)},\qquad   
\Gamma_X(q^2)=\Gamma_X D_X(q^2),
\end{eqnarray}
for $X=K^*$ or $K^*_0$. 
We adopt for the momentum-dependent widths the parametrizations of 
the $\tau$-decay program package TAUOLA \cite{Tauola}
\begin{eqnarray}
&&D_{K^*}(q^2)=\left\{\begin{array}{cl}
    \frac{m_{K^*}}{\sqrt{q^2}}
    \left[\frac{P_K(q^2)}{P_K(m^2_{K^*})}\right]^3& 
    \ \ {\rm for}\ \ q^2>(m_K+m_\pi)^2 \;,\\
    0 & \ \ {\rm for}\ \ q^2\leq (m_K+m_\pi)^2\;,
              \end{array}\right.
  \label{pi1_f}\\
&& { }\nonumber\\
&&D_{K^*_0}(q^2)=\left\{\begin{array}{cl}
    \frac{m_{K^*_0}}{\sqrt{q^2}}
    \left[\frac{P_K(q^2)}{P_K(m^2_{K^*_0})}\right]& 
    \ \ {\rm for}\ \ q^2>(m_K+m_\pi)^2 \;, \\
    0 & \ \ {\rm for}\ \ q^2\leq (m_K+m_\pi)^2\;.
              \end{array}\right.
  \label{pi2_f}
\end{eqnarray}

Let us now calculate the helicity amplitudes of the Cabbibo-suppressed
$\tau$ decay. In general, the $\tau$ three-body decay into $(K\pi)\nu_\tau$ 
is described by five independent kinematic variables. 
Because the hadronic $(K\pi)$ system is solely determined by the lepton 
momentum transfer, it is convenient to consider two reference frames, 
namely the $\tau$ rest frame and the $(K\pi)$ rest frame as shown in 
Fig.~\ref{fig:decay}. 
We define the momentum direction of the virtual $K^*$ or $K^*_0$ in the 
$\tau$ rest frame by the polar angle $\Theta$ and the azimuthal angle $\Phi$
with respect to the $\tau$ momentum direction taken to be along the positive
$z$-axis and the $K$ momentum direction in the $(K\pi)$ rest frame by the 
angles $\theta$ and $\phi$. The rotational invariance of the total system 
with respect to the $\tau$ momentum direction allows us to take $\Phi$ to be 
zero in the calculation of the helicity amplitudes, while the azimuthal angle
$\Phi$ is employed to describe $\tau$ polarization, especially the 
transverse polarization of the $\tau$. Setting $\Phi$ to be zero and using
the 2-component spinor technique \cite{Hagiwara1}, one can obtain the 
helicity amplitudes $M_\sigma$ of the Cabbibo-suppressed decay 
$\tau^-\rightarrow (K\pi)^-\nu_\tau$ in the reference frames defined in 
Fig.~\ref{fig:decay} as 
\begin{eqnarray}
M_+&=&2G_F\sin\theta_C(1+\chi)\sqrt{m_{\tau}^2 -q^2}
  \left[ m_{\tau}\cos\frac{\Theta}{2}
	\left(\frac{m^2_{K^*_0}}{q^2}+\xi\right)C_K F_S(q^2)
	\right. \nonumber\\
    && -2P_K \left. \left\{\cos\theta\cos\frac{\Theta}{2}
         +\sin\theta\sin\frac{\Theta}{2} e^{i\phi}
	+\left(\frac{m_{\tau}}{\sqrt{q^2}}-1\right)
	\cos\vartheta\cos\frac{\Theta}{2} \right\}F_K(q^2)
	\right],
  \nonumber\\
M_-&=&2G_F\sin\theta_C(1+\chi)\sqrt{m_{\tau}^2 -q^2}
  \left[
	m_{\tau}\sin\frac{\Theta}{2}
	\left(\frac{m^2_{K^*_0}}{q^2}+\xi\right)C_K F_S(q^2)
	\right.
	\nonumber\\
	&&+2P_K \left.
	\left\{\cos\theta\sin\frac{\Theta}{2}
         -\sin\theta\cos\frac{\Theta}{2} e^{-i\phi}
	 -\left(\frac{m_{\tau}}{\sqrt{q^2}}-1\right)
	\cos\vartheta\sin\frac{\Theta}{2}\right\}F_K(q^2)
	\right],
\label{mtamp} 
\end{eqnarray}
where $\vartheta$ is the angle between the $K$ momentum in the ($K\pi$) 
rest frame and the ($K\pi$) momentum in the $\tau$-rest frame, i.e. 
$\cos\vartheta = \sin\theta\cos\phi\sin\Theta +\cos\theta\cos\Theta$,
and $P_K$ is the size of the $K$ momentum in the virtual $K^*$ and $K^*_0$ 
rest frame;
\begin{eqnarray}
P_K(q^2)=\frac{1}{2\sqrt{q^2}}\lambda^{1/2}(q^2,m_{\pi}^2,m_K^2),
\end{eqnarray}
with $\lambda(x,y,z)=x^2+y^2+z^2-2xy-2yz-2xz$.
The dimensionless parameter $\xi$ determines the relative size of 
the scalar contributions to the vector ones: 
\begin{eqnarray}
\xi=\frac{m^2_{K^*_0}}{(m_s-m_u)m_\tau}
    \left(\frac{\eta}{1+\chi}\right).
\label{eq:xi}
\end{eqnarray}

Clearly, the initial $\tau^-$ system is not CP self-conjugate and so 
any genuine CP-odd observable can be constructed only by considering both the 
Cabbibo-suppressed $\tau^-$ decay and its charge-conjugated $\tau^+$ decay, 
and by identifying the CP relations of their kinematic distributions.
Before constructing all the possible CP-odd asymmetries explicitly, 
we calculate the helicity amplitudes for the charge-conjugated process
$\tau^+ \to (K\pi)^+ \bar{\nu}_\tau$.  
The helicity amplitudes $\overline{M}_{\bar{\sigma}}$ of the $\tau^+$ decay
in the same reference frame as in the $\tau^-$ decay are given by 
\begin{eqnarray}
\overline{M}_+&=&
2 G_F\sin\theta_C(1+\chi^*)\sqrt{m_{\tau}^2 -q^2}
  \left[
m_{\tau}\sin\frac{\bar{\Theta}}{2}
\left(\frac{m^2_{K^*_0}}{q^2}+\xi^*\right)C_K F_S(q^2)
\right. \nonumber\\
&&
-2P_K\left.\left\{
	-\cos\bar{\theta}\sin\frac{\bar{\Theta}}{2}
         +\sin\bar{\theta}\cos\frac{\bar{\Theta}}{2} e^{i\bar{\phi}}
	+\left(\frac{m_{\tau}}{\sqrt{q^2}}-1\right)
	\cos\bar{\vartheta}\sin\frac{\bar{\Theta}}{2}\right\}F_K(q^2)
	\right],
\nonumber\\ 
\overline{M}_-&=&-
2 G_F\sin\theta_C(1+\chi^*)\sqrt{m_{\tau}^2 -q^2}
  \left[  
m_{\tau}\cos\frac{\bar{\Theta}}{2}
\left(\frac{m^2_{K^*_0}}{q^2}+\xi^*\right)C_K F_S(q^2)
\right. \nonumber\\
&&
-2P_K\left.\left\{
        \cos\bar{\theta}\cos\frac{\bar{\Theta}}{2}
         +\sin\bar{\theta}\sin\frac{\bar{\Theta}}{2} e^{-i\bar{\phi}}
        +\left(\frac{m_{\tau}}{\sqrt{q^2}}-1\right)
        \cos\bar{\vartheta}\cos\frac{\bar{\Theta}}{2}\right\}F_K(q^2)
        \right].
\label{ptamp} 
\end{eqnarray}
It is easily shown that, if the parameters $\eta$ and $\chi$ are real,
the helicity amplitudes (\ref{mtamp}) for the $\tau^-$ decay and (\ref{ptamp})
for the $\tau^+$ decay satisfy the CP relation:
\begin{eqnarray}
               M_{\pm}(\Theta;q^2;\theta,\phi)
=\mp\overline{M}_{\mp}(\Theta;q^2;\theta,-\phi).
\label{cppro}
\end{eqnarray}

With the results in eqs.~(\ref{mtamp}) and (\ref{ptamp}) for 
the $\tau^\pm$ decay helicity amplitudes, one can
describe the decay of an arbitrary polarized $\tau$ lepton by superposing
the two helicity states. Generally, a pure $\tau$ polarization state, which is
polarized in the direction ($\theta_p, \phi_p$), is given in terms of 
the helicity states by
\begin{eqnarray}
|\theta_p,\;\phi_p\rangle=\cos\frac{\theta_p}{2}|+\rangle
     +\sin\frac{\theta_p}{2}e^{i\phi_p}|-\rangle\;,
\end{eqnarray}
where $\theta_p$ and $\phi_p$ are the polar and azimuthal angles 
in the $\tau$ rest frame, respectively, with respect to the $\tau$ 
momentum direction in the laboratory frame.
In this case the decay amplitude to the final state $|\Theta, \Phi\rangle$ of
the virtual $K^*/K^*_0$ system is expressed in terms of the helicity 
amplitudes as 
\begin{eqnarray}
\langle\Theta,\Phi|\theta_p,\phi_p\rangle
    =\langle\Theta,0|\theta_p,\phi_p-\Phi\rangle
    =\cos\frac{\theta_p}{2}M_+ +\sin\frac{\theta_p}{2}e^{i(\phi_p-\Phi)}M_-,
\end{eqnarray}
where the rotational invariance of the total system with respect to the
$\tau$ momentum direction has been used to relate the first and
second expressions. It is noteworthy that only the difference $\Phi-\phi_p$ 
between the azimuthal angles $\Phi$ and $\phi_p$ appears.

Taking into account the polarization degree $P$ for a 
partially-polarized $\tau^-$ beam, one can decompose the
differential decay rate into four independent terms: 
\begin{eqnarray}
d\Gamma&=&\frac{1}{2}d(\Gamma_{++}+\Gamma_{--})+
          \frac{1}{2}P\cos\theta_pd(\Gamma_{++}-\Gamma_{--})\nonumber\\
        &&+P\sin\theta_p\cos(\phi_p-\Phi) {\rm Re}(d\Gamma_{+-})
          -P\sin\theta_p\sin(\phi_p-\Phi) {\rm Im}(d\Gamma_{+-}),
\label{decayrate}
\end{eqnarray}
where the helicity-dependent terms are defined as
\begin{eqnarray}
d\Gamma_{\sigma\sigma^{\prime}}=
       \frac{1}{(2\pi)^5}\frac{1}{32m_{\tau}}
	\left(1-\frac{q^2}{m^2_{\tau}}\right)
       M_{\sigma}M^*_{\sigma^{\prime}}P_Kd\sqrt{q^2}\;
       d\cos\theta \;d\phi \;d\cos\Theta \;d\Phi.
\end{eqnarray}
For the sake of notational convenience, we introduce $d\Phi_3$ for the
phase space element $d\sqrt{q^2} \;d\cos\theta \;d\phi \; d\cos\Theta$ and 
denote the four independent terms in eq.~(\ref{decayrate}) as
\begin{eqnarray}
&&\frac{d\Gamma_1}{d\Phi_3}=\frac{d(\Gamma_{++}+\Gamma_{--})}{d\Phi_3}\,,\qquad
  \frac{d\Gamma_2}{d\Phi_3}=\frac{d(\Gamma_{++}-\Gamma_{--})}{d\Phi_3}\,,\nonumber\\
&&\frac{d\Gamma_3}{d\Phi_3}=2\;{\rm Re}\left(
                             \frac{d\Gamma_{+-}}{d\Phi_3}\right)\,,\qquad
  \frac{d\Gamma_4}{d\Phi_3}=2\;{\rm Im}\left(
                             \frac{d\Gamma_{+-}}{d\Phi_3}\right)\,.
\end{eqnarray}
Note that (a) the $\Gamma_1$ term is the unpolarized differential decay rate and
the other three terms  $\Gamma_i$ ($i=2,3,4$) are 
polarization-dependent; the $\Gamma_2$ term is due to longitudinal polarization,
the $\Gamma_{3}$ term due to transverse polarization and the $\Gamma_4$ 
due to normal polarization, and (b) the transverse and normal components 
after integrating the differential decay rate over the azimuthal angle
$\Theta$ or $\phi_p$ vanish, which is consistent with the so-called ``null 
transverse-polarization theorem" \cite{hikasa}. 
Clearly, to utilize the polarization-dependent
terms effectively, the initial $\tau$ beam should be highly polarized, 
that is to say, $P$ should be relatively large. Deferring the potential impact of 
the value of $P$ to our later discussion, we assume $P$=1 for the
time being. In this case, each $d\Gamma_i/d\Phi_3$ ($i=1$ to $4$) term 
can be decomposed into a CP-even part $\Sigma_i$  and a CP-odd part $\Delta_i$:
\begin{eqnarray}
\frac{d\Gamma_i}{d\Phi_3}=\frac{1}{2}(\Sigma_i+\Delta_i)\;.
\end{eqnarray}

The four CP-even parts $\Sigma_i$ and four CP-odd parts $\Delta_i$ can be easily
identified by use of the CP relation (\ref{cppro}) between the $\tau^-$ and $\tau^+$
decay helicity amplitudes and they are expressed in terms of the $\tau^\mp$ 
helicity-dependent terms $d\Gamma_i/d\Phi_3$ and $d\bar{\Gamma}_i/d\Phi_3$ as
\begin{eqnarray}
&&\Sigma_1=\frac{d(\Gamma_1+\bar{\Gamma}_1)}{d\Phi_3},\qquad
  \Sigma_2=\frac{d(\Gamma_2-\bar{\Gamma}_2)}{d\Phi_3},\nonumber\\
&&\Sigma_3=\frac{d(\Gamma_3-\bar{\Gamma}_3)}{d\Phi_3},\qquad
  \Sigma_4=\frac{d(\Gamma_4+\bar{\Gamma}_4)}{d\Phi_3},\nonumber\\
&&\Delta_1=\frac{d(\Gamma_1-\bar{\Gamma}_1)}{d\Phi_3},\qquad
  \Delta_2=\frac{d(\Gamma_2+\bar{\Gamma}_2)}{d\Phi_3},\nonumber\\
&&\Delta_3=\frac{d(\Gamma_3+\bar{\Gamma}_3)}{d\Phi_3},\qquad
  \Delta_4=\frac{d(\Gamma_4-\bar{\Gamma}_4)}{d\Phi_3},  
\label{cpcom}
\end{eqnarray}
where we have used the same kinematic variables $\{q^2, \Theta,\theta\}$
for the $d\bar{\Gamma}_i/d\Phi_3$ except for
the replacement of $\bar{\phi}$ by $-\phi$. We have numerically estimated 
the four CP-even terms in the allowed phase-space points and have found 
that the other three $\Sigma_i$ ($i=2,3,4$) terms are negligible compared to 
the $\Sigma_1$ term. Therefore we neglect those terms in the following.
The CP-even $\Sigma_1$ term and the CP-odd $\Delta_i$ ($i=1$ to $4$) can 
be obtained from the $\tau^\mp$ decay helicity amplitudes and their 
explicit form is listed in Appendix~A. All the CP-odd terms are proportional
to the imaginary part of the parameter $\xi$ in eq.~(\ref{eq:xi}).

An appropriate real weight function $w_i(\Theta;q^2,\theta,\phi)$
is usually employed to separate the $\Delta_i$ contribution and to enhance
its analysis power for the CP-odd parameter ${\rm Im}(\xi)$ through 
the CP-odd quantity:
\begin{eqnarray}
\langle w_i\Delta_i\rangle\equiv\int\left[w_i\Delta_i\right] d\Phi_3.
\end{eqnarray}
of which the analysis power is determined by the parameter 
\begin{eqnarray}
\varepsilon_i
   =\frac{\langle w_i\Delta_i\rangle}{\sqrt{\langle\Sigma_1\rangle
          \langle w_i^2\Sigma_1\rangle}}\;.
\label{Significance}
\end{eqnarray}
For the analysis power $\varepsilon_i$, the number $N_i$ of the $\tau$ leptons 
needed to observe CP violation at the 1-$\sigma$ level is
\begin{eqnarray}
N_i=\frac{1}{Br\cdot\varepsilon_i^2}\;,
\label{eq:number}
\end{eqnarray}
where $Br$ is the branching fraction of the relevant $\tau$ decay mode.
Certainly, it is desirable to find the optimal weight function
with the largest analysis power. It is known \cite{Optimal} that, 
when the CP-odd contribution to the total rate is relatively small, 
the optimal weight function  is approximately given by
\begin{eqnarray}
w^i_{\rm opt}=\frac{\Delta_i}{\Sigma_1}.
\end{eqnarray}
We adopt these optimal weight functions in the following numerical analyses
with several concrete models beyond the SM introduced in the following.

Although there is no CP violation in the $\tau$ decays within the SM,
it is possible to conceive several new sources of CP violation in the $\tau$ 
decays. Among them we will consider models with new scalar-fermion interactions,
which still preserve the SM gauge symmetries and have only the massless and 
left-handed neutrinos as in the SM. 
In this case, only four types of scalar-fermion interactions can 
contribute to the 
Cabbibo-suppressed decay  $\tau\rightarrow (K\pi)\nu_\tau$ \cite{Davies};
the MHD model \cite{Grossman} and three SLQ models \cite{Hall}. 

In the MHD model CP violation can arise in the charged Higgs sector
with more than two Higgs doublets \cite{Weinberg} and when not all the charged 
scalars are degenerate. As in most previous phenomenological analyses,
we also will assume in this MHD model that all but the lightest of the charged
scalars effectively decouple from fermions. The effective Lagrangian for the decay 
$\tau\rightarrow K\pi\nu_\tau$ in the assumption is then given at energies 
considerably low compared to $M_H$  by 
\begin{eqnarray}
{\cal L}_{\rm MHD}=2\sqrt{2}G_F\sin\theta_C
           \left(\frac{m_\tau m_s}{M^2_H}\right)
           \left[X^*Z(\bar{s}_R u_L)(\bar{\nu}_{\tau_L}\tau_R)
          +\left(\frac{m_u}{m_s}\right)Y^*Z
           (\bar{s}_L u_R)(\bar{\nu}_{\tau_L}\tau_R)\right]+{\rm h.c.},
\nonumber\\
\end{eqnarray}
where $X$, $Y$ and $Z$ are complex coupling constants which can be
expressed in terms of the charged Higgs mixing matrix elements.
From the effective Lagrangian, one obtain for the MHD CP-violation parameter
${\rm Im}(\xi_{\rm MHD})$ 
\begin{eqnarray}
{\rm Im}(\xi_{\rm MHD})=-\left(\frac{m_s}{m_s-m_u}\right)
                    \left(\frac{m^2_{K^*_0}}{M^2_H}\right)
    \left[{\rm Im}(XZ^*)+\left(\frac{m_u}{m_s}\right){\rm Im}(YZ^*)\right].
\label{MHDp}
\end{eqnarray}

On the other hand, the effective Lagrangians for the three SLQ models
\cite{Davies} contributing to the decay $\tau\rightarrow K\pi\nu_\tau$ 
are written in the form after a few Fierz rearrangements:
\begin{eqnarray}
&&{\cal L}^I_{\rm SLQ}=-\frac{x_{23}x^{\prime *}_{13}}{2M^2_{\phi_1}}
        \left[(\bar{s}_Lu_R)(\bar{\nu}_{\tau_L}\tau_R)
          +\frac{1}{4}(\bar{s}_L\sigma^{\mu\nu}u_R)
       (\bar{\nu}_{\tau_L}\sigma_{\mu\nu}\tau_R)\right] 
          +{\rm h.c.},\nonumber\\
&&{\cal L}^{II}_{\rm SLQ}=-\frac{y_{23}y^{\prime *}_{13}}{2M^2_{\phi_2}}
        \left[(\bar{s}_Lu_R)(\bar{\tau}^c_R\nu^c_{\tau_L})
          +\frac{1}{4}(\bar{s}_L\sigma^{\mu\nu}u_R)
           (\bar{\tau}^c_R\sigma_{\mu\nu}\nu^c_{\tau_L})\right]\nonumber\\
          &&\hskip 1.5cm +\frac{y_{23}y^*_{13}}{2M^2_{\phi_2}}
           (\bar{s}_L\gamma_\mu u_L)(\bar{\tau}^c_L\gamma^\mu\nu^c_{\tau_L})
            +{\rm h.c.},\nonumber\\
&&{\cal L}^{III}_{\rm SLQ}=-\frac{z_{23}z^*_{13}}{2M^2_{\phi_3}}
           (\bar{s}_L\gamma_\mu u_L)(\bar{\tau}^c_L\gamma^\mu\nu^c_{\tau_L})
            +{\rm h.c.}
\end{eqnarray}
Here the coupling constants $x^{(\prime)}_{ij}$, $y^{(\prime)}_{ij}$ and
$z_{ij}$  ($i,j=1,2,3$) are in general complex so that CP is violated
in the scalar-fermion Yukawa interaction terms. The superscript $c$ in the
Lagrangians ${\cal L}^{II}_{\rm SLQ}$ and ${\cal L}^{III}_{\rm SLQ}$ denotes 
charge conjugation, i.e. $\psi^c_{R,L}=i\gamma^0\gamma^2\bar{\psi}^T_{R,L}$ 
in the chiral representation. Although the tensor parts as well as the scalar 
parts appear in Model I and Model II, we do not have the tensor contributions
to $\tau\rightarrow (K\pi)\nu_\tau$ because we concentrate on the vector and 
scalar resonance contributions in the present work.
In the approximation that all the CP-even contributions from new interactions
are neglected, the size of the SLQ CP-violation effects is dictated by
the CP-odd parameters
\begin{eqnarray}
&&{\rm Im}(\xi^I_{\rm SLQ})=-\frac{m^2_{K^*_0}}{(m_s-m_u)m_\tau}
  \frac{{\rm Im}[x_{23}
      x^{\prime *}_{13}]}{4\sqrt{2}G_F\sin\theta_CM^2_{\phi_1}}\,,\nonumber\\
&&{\rm Im}(\xi^{II}_{\rm SLQ})=-\frac{m^2_{K^*_0}}{(m_s-m_u)m_\tau}
  \frac{{\rm Im}[y_{23}y^{\prime *}_{13}]}{4\sqrt{2}
         G_F\sin\theta_CM^2_{\phi_2}}\,,\nonumber\\
&&{\rm Im}(\xi^{III}_{\rm SLQ})=0\,.
\label{LQp}
\end{eqnarray}
This approximation is justified because the contributions from new physics 
are expected to be very small compared to those from the SM. 

The experimental constraints on the CP-violation parameters in 
(\ref{MHDp}) and (\ref{LQp}) depend on the values for $u$ and $s$ current 
quark masses, which are not well determined. Here, we use for the light $u$ 
and $s$ quark masses $m_u=5\ \ {\rm MeV}$ and $m_s=320\ \ {\rm MeV}$,
which satisfy the mass relation $m_s-m_u=7f_{K^*_0}$ \cite{Ayala}. 
Inserting the quark mass values into (\ref{MHDp}) and (\ref{LQp}) yields  
\begin{eqnarray}
{\rm Im}(\xi_{\rm MHD})&\simeq& -3.2\times 10^{-4}
      \left(\frac{M_W}{M_H}\right)^2
      \left[{\rm Im}(XZ^*)+\frac{1}{64}{\rm Im}(YZ^*)\right],\\
{\rm Im}(\xi^I_{\rm SLQ})&\simeq& -39.3
      \left(\frac{M_W}{M_{\phi_1}}\right)^2
      {\rm Im}[x_{23}x^{\prime *}_{13}]\;,\nonumber\\
{\rm Im}(\xi^{II}_{\rm SLQ})&\simeq& -39.3
      \left(\frac{M_W}{M_{\phi_2}}\right)^2
      {\rm Im}[y_{23}y^{\prime *}_{13}]\,,
\end{eqnarray}
where the $W$-boson mass $M_W$ is retained to show the $M_W$ dependence 
of the parameters explicitly, but 80 GeV \cite{PDG96} will be used for 
the $W$-boson mass in the actual numerical analysis.

The couplings $X$, $Y$ and $Z$ in the MHD model can be constrained through
the processes such as $B$-meson semileptonic decays. Because these experimental
constraints have been extensively reviewed in Ref.\ \cite{Grossman}, we simply
follow the analysis from which the combined constraint on 
${\rm Im}(\xi_{\rm MHD})$ is obtained to be
\begin{eqnarray}
|{\rm Im}(\xi_{\rm MHD})|<0.48,
\label{HDc}
\end{eqnarray}
when $M_H$ is set to be 45 GeV. Although there are at present no direct 
constraints on the SLQ CP-odd parameters in (\ref{LQp}), 
a rough constraint to the parameters can be provided on the assumption 
\cite{Davidson} that $|x^\prime_{13}|\sim |x_{13}|$ and 
$|y^\prime_{13}|\sim |y_{13}|$, that is to say, the leptoquark couplings 
to quarks and leptons belonging to the same generation are of a similar
size; then the experimental upper bound for the $D\bar{D}$ mixing yields
\begin{eqnarray}
|{\rm Im}(\xi^I_{\rm SLQ})|< 0.15,\qquad 
|{\rm Im}(\xi^{II}_{\rm SLQ})|< 0.14\;,
\label{ModelI}
\end{eqnarray}
which are stronger than the constraint (\ref{HDc}) on 
the MHD CP-odd parameter Im($\xi_{\rm MHD}$).
Based on the constraints (\ref{HDc}) and (\ref{ModelI}) to the CP-odd
parameters, we quantitatively estimate the number of the Cabbibo-suppressed
$\tau$ decays to detect CP violation for the maximally-allowed values 
of the CP-odd parameters:
\begin{eqnarray}
{\rm Im}(\xi_{\rm MHD})=0.48,\qquad
{\rm Im}(\xi^I_{\rm SLQ})=0.15,\qquad 
{\rm Im}(\xi^{II}_{\rm SLQ})=0.14.
\label{Bound}
\end{eqnarray}

As shown in eq.~(\ref{eq:number}) the branching fraction
for the Cabbibo-suppressed $\tau$ decay needs to be known beforehand, and
experimentally, it is crucial to
reconstruct the $\tau$ momentum direction. So we consider the most
easily-identifiable sequential decay channel $\tau\rightarrow (K^0\rightarrow 
\pi^-\pi^+)\pi^-\nu_\tau$ where an efficient identification of
the three charged-pion vertex and the $\tau^+\tau^-$ production vertex 
can be used to determine the momentum
of the mother $\tau$ lepton. The branching fraction for the clean sequential 
decay mode is approximately $0.33\%$ \cite{PDG96}. 
Table~1 shows the numbers $N_i$ of 
$\tau$ leptons required to detect CP violation through the CP-odd rate and 
polarization asymmetries $\Delta_i$ at the 1-$\sigma$ level for the 
maximally-allowed CP-odd MHD and SLQ parameters (\ref{Bound}) and $Br=0.33\%$.
All the CP-odd rate and polarization asymmetries require a similar number 
of $\tau$ decay events. But, we note that the results for the CP-odd 
polarization asymmetries have been obtained for  completely-polarized 
$\tau$ leptons. Therefore, in a realistic experiment with 
partially-polarized $\tau$ leptons the analysis power of the 
polarization-dependent observables will be reduced. 
In his recent works \cite{Tsai}, Tsai has claimed that $\tau$ polarization
can play a crucial role in probing P, CP and T violation in $\tau$ decays.
However, we see that at least in the Cabbibo-suppressed $\tau$ decay
$\tau\rightarrow (K\pi)\nu_\tau$ it is crucial to highly polarize the $\tau$
leptons to fully utilize the CP-odd polarization-dependent observables.  
Numerically, we find that CP violation in the MHD model and
the two SLQ models may be detected with about $10^6$ and $10^7$ $\tau$ leptons
for the maximally-allowed CP-odd MHD and SLQ parameters, respectively.

\vskip 1cm
\noindent
{Table~1}. {\it The number of $\tau$ leptons, 
$N_i$, needed for detection with the $\varepsilon_i$
at the $1\sigma$ level, are determined for $f_{K^*_0}=45 $ MeV
with ${\rm Im}(\xi_{\rm MHD})=0.48$ in the MHD model, and 
${\rm Im}(\xi^I_{\rm SLQ})=0.15$ and ${\rm Im}(\xi^{II}_{\rm SLQ})=0.14$
in the two SLQ models.} 
\vskip 0.8cm
\begin{center}
\begin{tabular}{|c|c|c|c|c|}\hline
\hskip 0.4cm Model\hskip 0.4cm { } 
& \hskip 1cm  $N_1$\hskip 1cm { }
& \hskip 1cm  $N_2$\hskip 1cm { }
& \hskip 1cm  $N_3$\hskip 1cm { } 
& \hskip 1cm  $N_4$\hskip 1cm { } \\
\hline
MHD    & $3.19\times 10^5$ & $3.20\times 10^5$ & $3.19\times 10^5$ & $3.21\times 10^5$\\
SLQI   & $3.26\times 10^6$ & $3.28\times 10^6$ & $3.27\times 10^6$ & $3.29\times 10^6$\\
SLQII  & $3.75\times 10^6$ & $3.76\times 10^6$ & $3.75\times 10^6$ & $3.78\times 10^6$\\ 
\hline
\end{tabular}
\end{center}
\vskip 1cm

In summary, we have investigated CP violation from the MHD model and
SLQ models in the Cabbibo-suppressed decay $\tau\rightarrow K\pi\nu_{\tau}$ 
with polarized $\tau$ leptons to which both the ${\rm J}^{\rm P}=1^-$ 
resonance $K^*$ and the ${\rm J}^{\rm P}=0^+$ resonance $K^*_0$ contribute. 
In addition to the CP-odd rate asymmetry, $\tau$ polarization enables us to
construct three additional CP-odd polarization asymmetries whose magnitudes 
depend crucially on the $K^*_0$ decay constant, $f_{K^*_0}$. 
Taking a QCD sum rule estimate of $f_{K^*_0}=45$ MeV and the present
experimental constraints on the CP-odd parameters into account, 
we have quantitatively estimated the maximally-allowed values for 
the CP-odd rate and polarization asymmetries in the MHD and SLQ models 
consistent with the SM gauge symmetry 
with massless left-handed neutrinos. We have found that the CP-odd rate and 
polarization asymmetries are of a similar size for highly-polarized $\tau$ 
leptons and, for their maximally-allowed values, new scalar-fermion interactions
may be detected with about $10^6$ or $10^7$ $\tau$ leptons at 
the $1\sigma$ level.
Consequently, we conclude that since the $\tau$ leptons of the order 
of $10^7$ or more
are expected to be produced yearly at the planned $B$ factories \cite{Bfactory} 
and the proposed $\tau$-charm factories \cite{taucharm},  it is
important to look for CP violation in the Cabbibo-suppressed 
$\tau$ decay $\tau \to K \pi \nu_\tau$ even without polarized
$\tau$ leptons.

\vskip 1cm

\section*{Acknowledgments.}

JS gratefully acknowledges support from the Korean Science
and Engineering Foundation (KOSEF) through the Center for Theoretical
Physics (CTP). The work was supported in part by the KOSEF-DFG
large collaboration project, Project No. 96-0702-01-01-2.

\vskip 2cm
\begin{center}
{\large\bf Appendix}
\end{center}

\appendix

\begin{appendix}

\subsection*{A.~CP-even and CP-odd observables}

In the appendix the explicit form of the CP-even term $\Sigma_1$ and
the CP-odd terms $\Delta_i$ ($i=1$ to $4$) is presented:
\begin{eqnarray}
\Sigma&=&F(q^2)\left[2C^2_K m^2_{\tau}
	\left|\frac{m^2_{K^*_0}}{q^2}+\xi\right|^2
     |F_S|^2+8P_K^2|F_K|^2
	\left\{1+\left(\frac{m^2_{\tau}}{q^2}-1\right)\cos^2\vartheta\right\}
     \right.\nonumber\\
     &&\left.\hskip 2cm -8C_KP_K\frac{m^2_{\tau}}{\sqrt{q^2}}
	\left(\frac{m^2_{K^*_0}}{q^2}
     +{\rm Re}(\xi)\right){\rm Re}(F_KF_S^*)\cos\vartheta\right]
\;,
\\
\Delta_1&=&-8F(q^2)C_Km_{\tau}P_K
	\left(\frac{m_{\tau}}{\sqrt{q^2}}\right)
         \cos\vartheta{\rm Im}(\xi){\rm Im}(F_KF^*_S)\;,\\
\Delta_2&=&-8F(q^2)C_Km_{\tau}P_K{\rm Im}(\xi)
	\left[\left\{\cos\theta+
       \left(\frac{m_{\tau}}{\sqrt{q^2}}-1\right)\cos\Theta\cos\vartheta\right\}
       {\rm Im}(F_KF^*_S)\right.\nonumber\\
       &&\mbox{ }\hskip 2cm 
         +\sin\Theta\sin\theta\sin\phi{\rm Re}(F_KF^*_S)\Bigg]\;,\\
\Delta_3&=&-8F(q^2)C_Km_{\tau}P_K{\rm Im}(\xi)
	\left[\left\{\sin\theta\cos\phi+
       \left(\frac{m_{\tau}}{\sqrt{q^2}}-1\right)\sin\Theta\cos\vartheta\right\}
       {\rm Im}(F_KF^*_S)\right.\nonumber\\
       &&\mbox{ }\hskip 2cm 
          -\cos\Theta\sin\theta\sin\phi{\rm Re}(F_KF^*_S)\Bigg]\;,\\
\Delta_4&=&-8F(q^2)C_Km_{\tau}P_K{\rm Im}(\xi)\bigg[\sin\theta\sin\phi
       {\rm Im}(F_KF^*_S)\nonumber\\
       &&\mbox{ } \hskip 2cm +(\cos\Theta\sin\theta\cos\phi
       -\sin\Theta\cos\theta){\rm Re}(F_KF^*_S)\bigg]\;,
\end{eqnarray}
where the overall function $F(q^2)$ is given by
\begin{eqnarray}
F(q^2)=\frac{G_F^2m_{\tau}\sin^2\theta_c}{2^7\pi^4}
      \left(1-\frac{q^2}{m_{\tau}^2}\right)^2|1+\xi|^2.
\end{eqnarray}

\end{appendix}

\vskip 2cm

\end{document}